\documentclass[epj]{svjour}
\usepackage{epsfig}
\usepackage{color}
\usepackage{amssymb}
\usepackage{amsmath, bbm}
\usepackage[figuresright]{rotating}

\usepackage[numbers,sort&compress]{natbib}
\usepackage{hyperref}

\newlength{\feynwidth} 

\newcommand{\be}{\begin{eqnarray}}
\newcommand{\ee}{\end{eqnarray}}

\begin{document}

\title{Structure of single-$\Lambda$ hypernuclei with chiral hyperon-nucleon potentials}

\author{Johann Haidenbauer$^1$ and Isaac Vida\~na$^2$}
\authorrunning{Johann Haidenbauer and Isaac Vida\~na}

\institute{
$^1$Institute for Advanced Simulation, Institut f\"{u}r Kernphysik (Theorie) and J\"{u}lich 
Center for Hadron Physics, Forschungszentrum J\"{u}lich, D-52425 J\"{u}lich, Germany \\
$^2$Istituto Nazionale di Fisica Nucleare, Sezione di Catania. Dipartimento di Fisica 
``Ettore Majorana", Universit\`a di Catania, Via Santa Sofia 64, I-95123 Catania, Italy
}
\date{Received: date / Revised version: date}
\abstract{
The structure of single-$\Lambda$ hypernuclei is studied using the chiral hyperon-nucleon potentials derived at leading order (LO) and next-to-leading order (NLO) by the J\"{u}lich--Bonn--Munich group. 
Results for the separation energies of $\Lambda$ single-particle states for various  hypernuclei from $^5_{\Lambda}$He to $^{209}_{\,\,\,\,\,\Lambda}$Pb are presented for the LO interaction and the 2013 (NLO13) and 2019 (NLO19) versions of the NLO potentials. 
It is found that the results based on the LO potential show a clear tendency for overbinding
while those for the NLO13 interaction underbind most of the considered hypernuclei.    
A qualitatively good agreement with the data is obtained for the NLO19 interaction over a fairly 
large range of mass number values when considering the uncertainty due to the regulator dependence. 
A small spin-orbit splitting of the $p$-, $d$-, $f$-, and $g$-wave states is predicted by all
interactions, in line with the rather small values observed in pertinent experiments. 
}

\maketitle
%%%%%%%%%%%%%%%%%%%%%%%%%%%%%%%%%%%%%%%%%%%%%%%%%%%%%%%%%%%%%%%%%%%%%%%%%%%%%%%%%%%%%%%%%%%%%%%%%%%%%%%%%%%%%%%%

\section{Introduction}

One of the goals of hypernuclear physics \cite{gal16,botta12,ht06} is to relate hypernuclei observables with the 
underlying bare hyperon-nucleon ($YN$) and hyperon-hyperon ($YY$) interactions \\
{
which, contrary to that between two nucleons ($NN$), are still poorly known due to 
the limited number and accuracy} of scattering data \cite{engelmann66,alexander68,sechi68,kadyk71}. 
Hypernuclei, therefore, constitute a valuable and complementary source of information to constrain better these interactions. 
In a simple picture, single-$\Lambda$ hypernuclei consist of an ordinary nucleus with the $\Lambda$ hyperon sitting in a single-particle state of an effective $\Lambda$-nucleus mean field potential. Based on this description, several approaches have been employed to study the properties of the $\Lambda$ hyperon in finite nuclei.  Woods-Saxon potentials, for instance, have been traditionally used to describe, in a shell model picture, the single-particle properties of the $\Lambda$ from medium to heavy hypernuclei \cite{bouyssy76,bouyssy79,dover80,motoba88}. Density dependent effects and non-localities have been included in non-relativistic Hartree-Fock calculations with Skyrme type $YN$ interactions in order to improve the overall fit of the single-particle energies \cite{skhf1,skhf2,skhf3,skhf4,skhf4b,skhf5,skhf5b,skhf5c,skhf5d}. The properties of hypernuclei have also been studied in a relativistic framework, such as Dirac phenomenology, where the hyperon-nucleus potential is derived from the nucleon-nucleus one \cite{dirac1,dirac2}, or 
from relativistic mean field theory 
\cite{rmf0,rmf1,rmf2,rmf3,rmf4,rmf5,rmf6,rmf7,rmf8,rmf8b,rmf9}. 
Microscopic hypernuclear structure calculations, which can provide the desired link between the hypernuclear observables and the bare $YN$ interaction, are also available. In these calculations the single-particle properties of the $\Lambda$ in the hypernucleus have  been mostly derived from effective $YN$ $G$-matrices built from bare $YN$ interactions \cite{g1,g1b,g1c,g1d,g2,g3,morten96,vidana98}. Recently, a quantum Monte Carlo calculation of single- and double-$\Lambda$ hypernuclei 
has been done using two- and three-body forces between the $\Lambda$ and the nucleons \cite{lonardoni1,qmchyp}. In most of these approaches, the quality of the description of hypernuclei relies on the validity of the mean field picture. However, the correlations induced by the $YN$ interaction can substantially change this picture and, therefore, should not be ignored a priori. Very recently, the spectral function of the $\Lambda$ in finite nuclei has been studied by one of the authors of the present work \cite{vidana17}, showing that the $\Lambda$ is less correlated than the nucleons in agreement with the idea that it maintains its identity inside the nucleus. The results of this study show also that in hypernuclear production reactions the $\Lambda$ hyperon is formed mostly in a quasi-free state.

Microscopic calculations of hypernuclear structure are usually based on realistic $YN$ interactions that describe the available scattering data in free space. These interactions have been mainly constructed within the framework of a 
meson-exchange theory \cite{juelich89,juelich04,nij1,nij2,nij3,nij4,nij5,nij6}. 
In recent years, however, $YN$ interactions have been also derived in SU(3) chiral effective field theory ($\chi$EFT) by the J\"{u}lich--Bonn--Munich group, first at leading order (LO) \cite{lo} and, then, at next-to-leading order (NLO) in the Weinberg power counting \cite{nlo13,nlo19}. The LO potential consists of four-baryon contact terms without derivatives and of one-pseudoscalar-meson exchanges whereas at NLO contact terms with two derivatives arise together with loop contributions from (irreducible) two-pseudoscalar-meson exchanges (see Fig. 1 of Ref. \cite{nlo13}). Corresponding and consistent hyperon-nucleon-nucleon ($YNN$) interactions have been likewise derived by the J\"{u}lich--Bonn--Munich group in $\chi$EFT \cite{nny}. 
The leading contributions, consisting of three-body forces (3BFs) due to two-meson exchange, 
one-meson exchange, and six-baryon contact terms, 
appear at next-to-next-to-leading order (N$^2$LO) in the applied Weinberg counting.  

In the present work we examine the chiral LO and NLO $YN$ interactions \cite{lo,nlo13,nlo19} in 
microscopic hypernuclear-struc\-ture calculations. So far, these $YN$ potentials from chiral EFT have been 
employed only in studies of light hypernuclei \cite{Nogga:2013,nlo19,Wirth:2014,Gazda:2015,Wirth:2016,Gazda:2016,Wirth:2019},
and of the properties of hyperons in (infinite) nuclear matter \cite{kohno10,haidenbauer15,petschauer16,haidenbauer17,kohno18} 
using the Brueckner--Hartree--Fock (BHF) approach. 
For instance, in Ref. \cite{haidenbauer17}, it was found that the $\Lambda$ single-particle potential 
at zero momentum in symmetric nuclear matter and pure neutron matter becomes repulsive for densities larger than about two times normal nuclear saturation density, in contrast with the results obtained in similar BHF calculations with more conventional meson-exchange potentials. From this result, the authors of Ref.\ \cite{haidenbauer17} concluded that in neutron star matter this repulsion would shift the onset of hyperons to high densities potentially solving the so-called ``hyperon puzzle", {\it i.e.,} the difficulty to reconcile the recent observation of 2$M_\odot$ neutron stars with the presence of hyperons in their interiors (see {\it e.g.,} Ref.\ \cite{debbie14} and references therein).

Faddeev-type calculations of the hypertriton, $^3_\Lambda$H, performed with the LO \cite{Nogga:2013} 
and NLO \cite{nlo19} interactions, yield a satisfactory value for the separation energy. 
However, one should keep in mind that this energy has been actually used as an additional 
constraint for fixing the relative strength of the singlet- and triplet 
$\Lambda N$ $S$-wave interactions \cite{lo,nlo13,nlo19}. 
Calculations of the separation energies for the four-body hypernuclei ${}^4_\Lambda {\rm H}$ 
and ${}^4_\Lambda {\rm He}$ based on the EFT interactions revealed that there is only a
qualitative agreement with the experiments \cite{Nogga:2013,nlo19}. 
Specifically, the $0^+$ state is slightly overbound by the LO interaction \cite{Nogga:2013}
and noticeably underbound by the NLO interactions \cite{nlo19}. For fairness one has to say 
that an underbinding is likewise observed for phenomenological models of the $YN$ 
interaction \cite{nlo19}. 
The LO interaction has been also used in studies of light hypernuclei from ${}^4_\Lambda {\rm H}$/${}^4_\Lambda {\rm He}$ 
to $^{13}_{\,\,\, \Lambda}$C in {\it ab initio} calculations based on the no-core shell model (NCSM) 
\cite{Wirth:2014,Gazda:2015,Wirth:2016,Gazda:2016,Wirth:2019} and corresponding results for the NLO interaction
are becoming available too \cite{Nogga:2019,Le:2019}. 

The manuscript is organized in the following way. In Sec.\ \ref{sec:sec2} we  briefly describe the method 
to obtain the $\Lambda$ single-particle properties in finite nuclei. Results for 
a variety of single-$\Lambda$ hypernuclei are shown in Sec.\ \ref{sec:sec3}. Finally, a brief summary and some concluding remarks 
are given in Sec.\ \ref{sec:sec4}.

\section{$\Lambda$ single-particle properties in finite nuclei}
\label{sec:sec2}

The aim of this work is to study the properties of the $\Lambda$ in several nuclei using the chiral $YN$ interactions of the J\"{u}lich--Bonn--Munich group at LO and NLO. 
We follow a perturbative many-body approach to calculate the $\Lambda$ self-energy in finite nuclei from which we can determine the different $\Lambda$ 
single-particle bound states in the corresponding hypernuclei under study.  
This approach was originally developed to study the properties of the nucleon \cite{borromeo92} and the $\Delta$ \cite{morten94} isobar in finite nuclei, and has been already extended to study those of the $\Lambda$ and $\Sigma$ hyperons in Refs.\ \cite{morten96,vidana98,vidana17} using meson-exchange $YN$ interactions. In the following we present a brief summary of this approach and refer the interested reader to Refs.\ \cite{borromeo92,morten94,morten96,vidana98,vidana17} for a detailed description.  The approach starts with the construction of all the $YN$ $G$-matrices which describe the interaction between a hyperon (Y=$\Lambda$, $\Sigma$) and a nucleon in infinite nuclear matter. 
To such end the coupled-channel Bethe--Goldstone equation is solved in momentum space including partial waves up to a maximum value of the total angular momentum $J=4$. We note here that, when solving it, the so-called discontinuous prescription has been adopted. These $G$-matrices are then used to obtain the $YN$ $G$-matrices in finite nuclei through the following integral equation:
\begin{eqnarray}
G_{FN}&=&G+G\left[\left(\frac{Q}{E}\right)_{FN}-\left(\frac{Q}{E}\right)\right]G_{FN} \nonumber \\
&=&G+G\left[\left(\frac{Q}{E}\right)_{FN}-\left(\frac{Q}{E}\right)\right]G \nonumber \\
&+&G\left[\left(\frac{Q}{E}\right)_{FN}-\left(\frac{Q}{E}\right)\right]G\left[\left(\frac{Q}{E}\right)_{FN}-\left(\frac{Q}{E}\right)\right]G \nonumber \\
&+& \cdot \cdot \cdot \ ,
\label{eq:gfng}
\end{eqnarray}
which expresses the finite nuclei $G$-matrices, $G_{FN}$, in terms of the nuclear matter ones, $G$, and the difference 
between the finite-nucleus and the nuclear-matter propagators, written schematically as $(Q/E)_{FN}-(Q/E)$. This difference, which accounts for the relevant 
intermediate particle-particle states has been shown to be quite small \cite{borromeo92,morten94,morten96,vidana98,vidana17} and, thus, 
in all practical calculations $G_{FN}$ can be well approximated by truncating the expansion (\ref{eq:gfng}) up to second order in 
the nuclear matter $G$-matrices. Therefore, we have
\begin{equation}
G_{FN} \approx G+G\left[\left(\frac{Q}{E}\right)_{FN}-\left(\frac{Q}{E}\right)\right]G \ .
\label{eq:g2nd}
\end{equation}

\begin{figure}[t!]
\begin{center}
\includegraphics[width=8.5cm,angle=-00,clip]{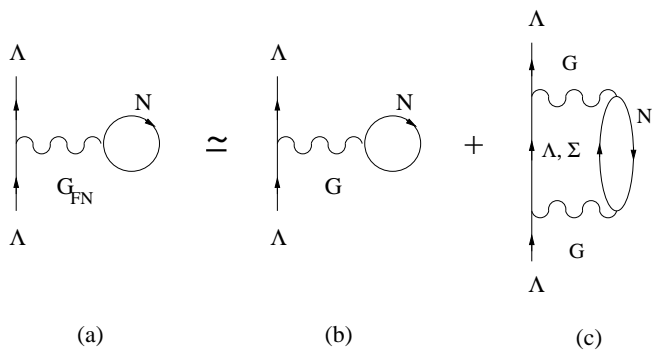}
\caption{BHF approximation of the finite nucleus $\Lambda$ self-energy (diagram (a)), split into the sum of a first-order contribution (diagram (b)) and a second order
2p1h correction (diagram (c)).}
\label{fig:self}
\end{center}
\end{figure}

Using then $G_{FN}$ as an effective $YN$ interaction we obtain the $\Lambda$ self-energy in the BHF approximation (see diagram (a) of Fig.\ \ref{fig:self}). This approximation can be split into the sum of two contributions: the one shown by diagram (b), which originates from the first-order term on the right-hand side of Eq.\ (\ref{eq:g2nd}), and that of diagram (c), which stands for the so-called {\it two-particle-one-hole} (2p1h) correction,  where the intermediate particle-particle propagator has to be viewed as the difference of propagators 
appearing in Eq.\ (\ref{eq:g2nd}). Solving the Sch\"{o}dinger equation with the real part of the $\Lambda$ self-energy we are able to determine, as we just mentioned, the  different $\Lambda$ single-particle bound states. 

Before we present and analyze our results in detail, we would like to point out a particular feature of 
the employed method that will be of relevance below. 
The distortion of the plane wave associated with the nucleon in the intermediate state of the 2p1h diagram of Fig.\ \ref{fig:self}(c), necessary to ensure 
its orthogonalization to the nucleon hole states, is taken into account only approximately. The orthogonalization procedure, described in detail in 
Ref.\ \cite{borromeo92}, was originally optimized for the case of $^{17}_{\ \Lambda}$O \cite{morten96}. As a consequence, the method tends to overestimate the binding
energies for very heavy hypernuclei such as $^{91}_{\ \Lambda}$Zr and $^{209}_{\ \Lambda}$Pb. 
Indeed, as discussed already in Refs.\ \cite{vidana98,vidana17}, and as we will see below, 
in general it leads to results where the $\Lambda$ is more bound in those nuclei than in infinite nuclear matter. 
This, of course, represents a clear signal for limitations in the applicability of that method to the heaviest hypernuclei.

%%%%%%%%%%%%%%%%%%%%%%%%%%%%%%%%%%%%%%%%%%%%%%%%%%%%%%%%%%%%%%%
%%%%%%%%%%%%%%%%%%%%%%%%%%%%%%%%%%%%%%%%%%%%%%%%%%%%%%%%%%%%

\section{Results and discussion}
\label{sec:sec3}

%%%%%%%%%%%%%%%%%%%%%%%%%%%%%%%%%%%%%%%%%%%%%%%%%%%%%%%%%

\begin{table*}[t]
\begin{center}
\scriptsize
\renewcommand{\arraystretch}{1.12}
\begin{tabular}{c|cccc|ccccc|ccccc|c}
\hline
\hline
 & & \multicolumn{3}{c}{LO} & \multicolumn{5}{c}{NLO13} & \multicolumn{5}{c}{NLO19} & \multicolumn{1}{c}{Exp.} \cr
\hline 
{$\Lambda$ (MeV)} & $550$ &$600$ &$650 $&$700$ & $500$ & $550$ & $600$ & $650$ & $700$ & $500$ & $550$ & $600$ & $650$ & $700$ & \\
\hline  
 & & &  & & &  &  &  &  &  &   &  & &  \\
$^5_{\Lambda}$He & & & & &  & & &  &  &  &  &  &   &  & $^5_{\Lambda}$He \\
 $s_{1/2}$  & $4.04$& $3.32$ & $3.06$ & $3.26$ & $0.73$ & $0.15$& $0.63$ & $2.36$ &  $4.90$ &  $2.16$& $1.36$ & $1.77$ & $3.42$  & $5.63$ & $3.12(2)$  \\
 & & &  & & &  &  &  &  &  &  &  & & &  \\
\hline
& & &  & & &  &  &  &  &  &  &   & & &  \\
$^{13}_{\,\,\,\Lambda}$C & & &  & & &  & & &  &  &  &  &   &  & $^{13}_{\,\,\,\Lambda}$C \\
 $s_{1/2}$  & $12.33$ & $11.01$& $10.54$ & $10.93$ & $4.44$ & $2.24$ & $3.72$& $8.91$ & $13.40$ &  $8.91$ & $6.42$& $7.22$ & $10.81$ & $14.98$ & $11.69(12)$ \\
 $p_{3/2}$   &  $-$ &  $-$  & $-$  &  $-$  & $-$  &  $-$       &  $-$  & $-$ & $1.22$ & $-$ &  $-$  & $-$  &  $0.12$ & $1.76$ & $0.8(3)$ (p) \\
 $p_{1/2}$   &  $1.11$ &  $0.58$  & $0.45$  &  $0.72$ & $-$  &  $-$   &  $-$  & $-$ & $0.97$ & $-$ &  $-$  & $-$  &  $-$ & $1.40$& \\
 & & &  & & &  &  &  &  &  &  & &  &  \\
\hline
 & & &  & & &  &  & & &  &  &  &  &   &  \\
$^{17}_{\,\,\,\Lambda}$O  & & & & &  & & &  &  &  &  &  &   &  & $^{16}_{\,\,\,\Lambda}$O \\
 $s_{1/2}$  & $16.12$ & $14.64$& $14.13$ & $14.65$  & $6.07$ & $3.46$ & $5.35$& $10.51$ & $16.37$ & $11.46$ & $8.61$& $9.55$ & $13.60$  & $18.18$ & $13.0(2)$ \\
 $p_{3/2}$   &  $3.16$  &  $2.29$  & $2.02$  &  $2.30$  & $-$  &  $-$       &  $-$  & $1.22$  & $4.04$ & $1.26$  &  $0.14$  & $0.53$  &  $2.40$  & $4.89$ & $2.5(2)$ (p) \\
 $p_{1/2}$   &  $3.47$  &  $2.64$  & $2.41$  &  $2.76$   & $-$  &  $-$       &  $-$  & $0.66$ & $3.31$ & $0.51$  &  $-$  & $-$  & $1.69$  & $4.10$ & \\
 & & & & &  & & &  &  &  &  &  &  &  &  \\
\hline
 & & & & &  & & &  &  &  &  &  &  &   &  \\
$^{41}_{\,\,\,\Lambda}$Ca & & & & &  & & &  &  &  &  &  &  &   & $^{40}_{\,\,\,\Lambda}$Ca \\
 $s_{1/2}$  & $24.83$  & $23.17$ & $22.66$  &$23.26$  & $12.37$ & $8.78$  & $11.24$ & $17.56$ & $24.36$ &   $19.51$  & $15.86$ & $16.80$  & $21.30$  & $26.47$  
& $18.7(1.1)^\dagger$   \\ 
 $p_{3/2}$  & $14.50$   & $13.05$ & $12.54$   &$12.95$  & $4.95$ & $2.54$   & $3.98$ & $8.82$  &$13.43$ & $9.91$   & $6.93$ & $7.48$   &$11.04$  & $15.06$ & $11.0(5)$ (p) \\
 $p_{1/2}$  & $14.70$   & $13.28$   & $12.81$    &  $13.25$  & $4.37$ & $2.08$   & $3.50$   &$7.73$  & $12.87$  & $9.13$   & $6.23$   & $6.82$   & $10.42$ &  $14.47$ & \\
 $ d_{5/2}$  & $4.61$    & $3.45$   &$3.01$     &$3.23$   & $-$ & $-$  & $-$   &$0.40$  & $3.59$   & $1.47$    & $-$   &$-$     &$1.99$  & $4.67$   & $1.0(5)$ (d) \\
 $d_{3/2}$  & $6.91$     & $5.64$   &$5.18$     & $5.51$  & $-$ & $-$            & $-$   &$0.50$  & $4.02$   & $0.56$     & $-$   &$-$     & $1.20$ & $3.84$ &  \\
 & & & & &  & & &  &  &  &  &  &  &  &   \\
\hline
& & & & &  & & &  &  &  &  &  &  &   &  \\
$^{91}_{\,\,\,\Lambda}$Zr & & & & &  & & &  &  &  &  &  &   & & $^{89}_{\,\,\,\Lambda}$Y \\
$s_{1/2}$  &$31.27$ &$29.22$ &$28.48$ & $29.11$   & $19.36$ &$14.66$ &$17.83$ &$25.10$  & $32.50$ & $27.72$ &$22.57$ &$23.19$ & $28.94$   & $34.61$ & $23.6(5)$ \\ 
$p_{3/2}$  &$24.31$ &$22.43$ &$21.72$ &$22.22$   & $14.24$ &$10.59$ &$13.27$ &$19.27$  & $25.45$ & $20.59$ &$16.24$ &$16.94$ & $22.05$   & $26.96$ & $17.7(6)$ (p) \\
$p_{1/2}$  &$24.80$ &$22.96$ &$22.28$ &$22.80$   &$13.95$ &$10.39$ &$13.05$ &$19.07$  & $25.31$ & $20.45$ &$15.96$ &$16.67$ &$21.86$  & $26.82$ & \\
$d_{5/2}$ &$16.60$ &$14.79$  &$14.09$ &$14.30$   &$6.21$ &$3.33$ &$5.24$  & $10.30$ & $15.27$ & $11.92$ &$8.10$  &$8.44$ &$12.68$   & $16.78$ & $10.9(6)$ (d) \\
$d_{3/2}$  &$17.57$ &$15.80$  &$15.06$ &$15.40$   &$5.80$ &$2.98$ &$4.88$  & $9.70$  & $14.97$ & $11.65$ &$7.61$  &$7.98$ &$12.27$   & $16.40$ &  \\
$f_{7/2}$  &$8.69$ &$7.04$   &$6.25$ &$6.36$           &$-$ &$-$ &$-$   &$1.68$    & $5.63$  & $4.04$ &$0.98$   &$0.89$ &$3.97$       & $7.04$ & $3.7(6)$ (f) \\
$f_{5/2}$  &$10.17$ &$8.58$   &$7.85$ &$8.04$          &$-$ &$-$ &$-$   &$1.28$  & $5.23$   & $3.59$ &$0.33$   &$0.28$ &$3.39$       & $6.54$ & \\
& & & & &  & & &  &  &  &  &  &  &   &  \\
\hline
& & & & &  & & &  &  &  &  &  &  &  &   \\
$^{209}_{\,\,\,\,\,\Lambda}$Pb & & & & &  & & &  &  &  &  &  &  &   & $^{208}_{\,\,\,\,\,\Lambda}$Pb \\
$s_{1/2}$   &$40.97$ &$38.18$ &$36.91$ &$37.32$  &$25.75$ &$21.41$ &$25.09$ &$32.28$  & $39.51$ & $36.28$ &$29.50$ &$29.60$ & $35.84$  & $41.58$ & $26.9(8)$ \\ 
$p_{3/2}$  &$37.62$ &$34.85$ &$33.42$ &$33.50$  &$21.88$ &$15.77$ &$18.33$ &$25.13$ & $31.83$  & $33.72$ &$26.73$ &$25.27$ & $30.26$ & $34.71$ & $22.5(6)$ (p) \\
$p_{1/2}$  &$37.81$ &$35.05$ &$33.62$ &$33.69$  &$21.55$ &$15.53$ &$18.14$ &$25.00$  & $31.74$ & $33.58$ &$26.57$ &$25.13$ & $30.17$ & $34.64$  & \\
$d_{5/2}$  &$29.57$ &$26.94$ &$25.45$ &$25.29$  &$14.47$ &$8.79$ &$9.96$ &$14.78$  & $19.98$ & $25.49$ &$19.28$ &$16.84$ & $20.08$ & $23.15$ & $17.4(7)$ (d) \\
$d_{3/2}$  &$30.03$ &$27.44$ &$26.00$ &$25.87$  &$14.35$ &$8.71$ &$9.83$ &$14.62$ & $19.83$ & $25.29$ &$18.98$ &$16.57$ & $19.85$ & $22.97$ & \\
$f_{7/2}$  &$22.85$ &$20.33$ &$18.92$ &$18.79$   &$4.46$ &$-$ &$-$ &$5.91$  & $12.57$ & $16.23$ &$10.15$ &$7.91$ &$11.90$  & $15.80$ & $12.3(6)$ (f) \\
$f_{5/2}$  &$23.45$ &$20.95$ &$19.57$ &$19.48$   &$4.42$ &$-$ &$-$ &$5.60$ & $12.24$ & $15.96$ &$9.70$ &$7.47$ &$11.47$ & $15.38$ & \\
$g_{9/2}$  &$20.33$ &$17.75$ &$16.23$ &$15.96$  &$1.87$ &$-$ &$-$ &$3.23$  & $9.21$ & $13.72$ &$7.55$ &$5.18$ &$8.92$ & $12.32$ & $7.2(6)$ (g) \\
$g_{7/2}$  &$21.47$ &$18.96$ &$17.50$ &$17.29$   &$1.38$ &$-$ &$-$ &$2.91$ & $8.94$ & $13.38$ &$7.03$ &$4.69$ &$8.53$ & $12.00$ & \\
& & &  & & &  &  &  & &  &  &  &  &   \\
\hline
\hline
$-U_{\Lambda}(k=0)$ &$38.75$ &$37.31$ &$36.22$ &$35.62$  &$31.47$ &$26.20$ &$25.22$ &$27.44$  & $28.50$ & $40.99$ &$38.67$ &$35.43$ & $33.76$  & $32.25$ & $27-30$ \\
\hline
\hline
\end{tabular}
\renewcommand{\arraystretch}{1.0}
\end{center}
\caption{Separation energies of $\Lambda$ single-particle states (in MeV) of several hypernuclei from $^5_{\Lambda}$He to 
$^{209}_{\,\,\,\,\,\Lambda}$Pb, and the BHF $\Lambda$ single-particle potential at zero momentum, -$U_{\Lambda}(k=0)$, 
in symmetric nuclear matter at saturation density ($\rho_0=0.17$ fm$^{-3}$). 
Results are 
shown for the chiral $YN$ interactions at LO \cite{lo} and NLO \cite{nlo13,nlo19} for different values of the cutoff $\Lambda$. 
Available experimental data \cite{gal16,Botta:2017} for the closest measured hypernuclei are included. 
$^\dagger$The weak signal for $^{40}_{\,\,\,\Lambda}$Ca \cite{Pile:1991} is not included in the recent compilation \cite{gal16}. 
 }
\label{tab:sp}
\end{table*}

%%%%%%%%%%%%%%%%%%%%%%%%%%%%%%%%%%%%%%%%%%%%%%%%%%%%%%%%%

The separation energies of the different $\Lambda$ single-particle states in $^5_\Lambda$He, $^{13}_\Lambda$C, $^{17}_\Lambda$O, 
$^{41}_\Lambda$Ca, $^{91}_\Lambda$Zr and $^{209}_{\,\,\,\,\,\Lambda}$Pb obtained with the chiral interactions are 
summarized in Table~\ref{tab:sp}. 
Note that all hypernuclei considered in the present work consist of a closed-shell nuclear core plus a $\Lambda$ 
sitting in a single-particle state. 
The values reported are to be compared with the experimental separation energies for the corresponding hypernuclei. 
However, since experimental data for the particular hypernuclei we consider do not always exist, 
we include for comparison the closest representative hypernuclei for which experimental information is 
available. However, one should not attribute the differences between the calculated and the experimental 
values to this fact. These are mainly due to the approximative character of the calculation and/or, of course, 
due to the properties of the employed $YN$ interactions. 

Results are presented for the LO interaction from 2006 \cite{lo} and the two NLO versions from 2013 (NLO13) 
\cite{nlo13} and 2019 (NLO19) \cite{nlo19}. The latter interactions differ by different choices for the 
low-energy constants (LECs) that determine the strength of the contact interactions.   
In the initial NLO potential published in 2013 \cite{nlo13} all LECs in the $S$-waves were fixed by a 
fit to the available $\Lambda p$ and $\Sigma N$ data at low energies. 
In the NLO19 potential some of the $S$-wave LECS were inferred from the $NN$ sector via the underlying 
SU(3) flavor symmetry so that only a reduced number of LECs needed to be determined from the empirical 
information in the $YN$ sector, see Ref.~\cite{nlo19} for details. 
The interaction in the $P$-waves and in higher partial waves is the same in the two potentials. 
As discussed thoroughly in Ref.~\cite{nlo19} the results for $\Lambda p$ and $\Sigma N$ scattering
observables for the NLO13 and NLO19 potentials are practically identical. 
However, there is a considerable difference in the strength of the $\Lambda N \to \Sigma N$
transition potential between the two interactions and that has an influence on few- and many-body
applications like the one in the present work. 

Chiral baryon-baryon interactions derived within the Weinberg scheme require a regularization when 
inserted into the scattering (Lippmann-Schwinger) or $G$-matrix \\ (Bethe-Goldstone) equations
\cite{Epelbaum:2009}. 
As a consequence, the results depend on the regulator, specifically on the cutoff mass $\Lambda$ in the 
exponential regulator function adopted in the chiral $YN$ potentials of the 
J\"ulich--Bonn--Munich group \cite{lo,nlo13,nlo19}. 
Since, in principle, observables should not depend on the regulator, the actual regulator 
dependence provides arguably a measure for the magnitude of (missing) higher order contributions 
to the $YN$ interaction, and in case of applications to few- and many-body systems also for missing 
many-body forces, in particular of 3BFs \cite{Nogga:2013,Hammer:2013}. 
Accordingly, the results for the chiral interactions in Table~\ref{tab:sp} are given for a range of 
cutoff masses. The variation with the cutoff has to be considered as a lower bound for the uncertainty 
due to truncation of the chiral expansion.
The range chosen for $\Lambda$ is similar to what was used for chiral $NN$ potentials \cite{Epelbaum:2009}. 
The actual values, $\Lambda=550$--$700$ MeV (for the LO interaction) and
$\Lambda=500$--$650$ MeV (NLO), correspond to the range where the best description (lowest $\chi^2$)
of $\Lambda p$ and $\Sigma N$ scattering data was achieved \cite{nlo13,nlo19}.
NLO results for the cutoff $\Lambda=700$ MeV are included here for reasons discussed below.

We want to emphasize that no modifications of the underlying $YN$ interactions
are applied in the course of our calculation
to improve the description of hypernuclei.

Let us analyze the results now. Interestingly, there is a rather good agreement between experimental data 
and the results obtained with the LO chiral $YN$ interaction for $^5_\Lambda$He and $^{13}_\Lambda$C in the 
whole range of cutoff values. However, it is informative to compare our predictions for 
these two hypernuclei with the ones of a recent {\it ab initio} calculation based on the NCSM 
by Wirth and Roth \cite{Wirth:2019} where likewise this LO $YN$ potential was employed. 
In principle, the latter calculation should yield ``exact'' results when fully converged. 
(NCSM results for $^4_{\Lambda}$He \cite{Wirth:2016,Gazda:2016} based on the LO interaction agree 
within $200$ keV with those from a Faddeev-Yakubovsky calculation \cite{Nogga:2013}.)
In Ref.~\cite{Wirth:2019} a separation energy of around $4.4-4.7$ MeV was reported for 
$^5_{\Lambda}$He and $13.5 - 14.5$ MeV for $^{13}_{\,\,\, \Lambda}$C, 
for the LO interaction with cutoff $\Lambda = 700$ MeV. These values differ from our predictions 
by about $1$ MeV for the former and by roughly $3$~MeV for the latter hypernucleus, cf. Table~\ref{tab:sp}. 
Thus, taking the results of Wirth and Roth~\cite{Wirth:2019} as guideline for what should be the 
``true'' values for the LO interaction, one can conclude that the approach we follow could 
underestimate the actual values of the binding energies of light hypernuclei by $20$ to $30$\,\%. 
In any case, the $\Lambda$ single-particle energies for the other hypernuclei calculated with the LO potential, 
specially for the heavier 
ones $^{91}_\Lambda$Zr and $^{209}_{\,\,\,\,\,\Lambda}$Pb, appear clearly overbound with respect to 
the experimental values, cf. Table \ref{tab:sp}. This overbinding is also observed in 
the BHF result of the $\Lambda$ single-particle potential in symmetric nuclear matter, $U_{\Lambda}(k=0)$, 
at the saturation density of $\rho_0=0.17$ fm$^{-3}$ which is predicted to be deeper than expected from 
the analysis of hypernuclear data, see last line of Table~\ref{tab:sp}, 
indicating that the $YN$ $G$-matrices obtained with chiral interactions at LO are too attractive.

%%%%%%%%%%%%%%%%%%%%%%%%%%%%%%%%%%%%%%%%%%%%%%%%%%%%%%%%%%%%%%%%%%%%%%%%%%%%%%%%%%%%%%%%%%%%%%%%%%%%%%%%%%%
\begin{figure}[t]
\begin{center}
\includegraphics[width=8.5cm]{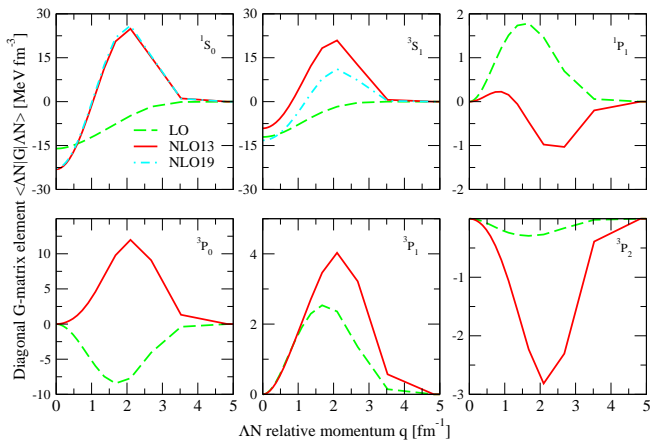}
\caption{$S$- and $P$-wave diagonal $\Lambda N$$\rightarrow$$\Lambda N$ nuclear matter $G$-matrices as a function 
of the $\Lambda N$ relative momentum. Results are shown for the LO (dashed line), NLO13 (solid line) and 
NLO19 (dash-dotted line) potentials with a cutoff of 600 MeV. 
The potentials NLO13 and NLO19 yield practically identical $G$-matrices for $P$- and higher order waves.}
\label{fig:matel}
\end{center}
\end{figure}
%%%%%%%%%%%%%%%%%%%%%%%%%%%%%%%%%%%%%%%%%%%%%%%%%%%%%%%%%%%%%%%%%%%%%%%%%%%%%%%%%%%%%%%%%%%%%%%%%%%%%%%%%%%
 
%%%%%%%%%%%%%%%%%%%%%%%%%%%%%%%%%%%%%%%%%%%%%%%%%%%%%%%%%%%%%%%%%%%%%%
\begin{figure}[t!]
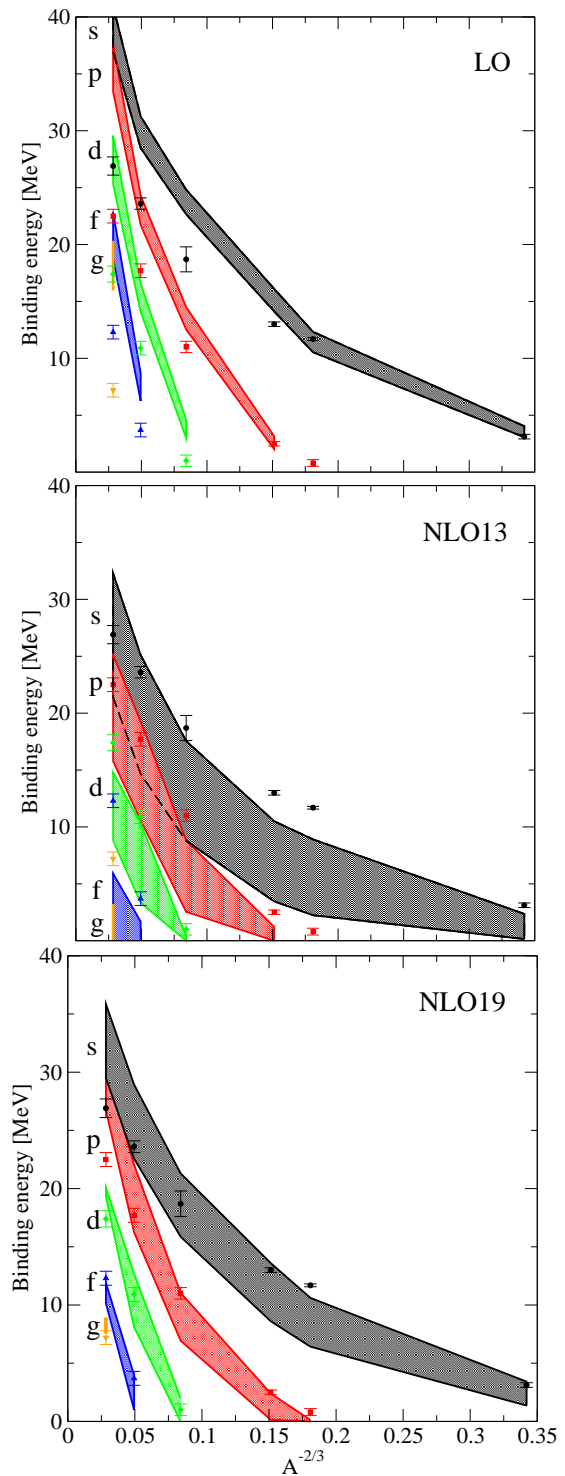

\begin{center}
\vskip -0.cm
\includegraphics[width=6.8cm,angle=0,clip]{fig3a.eps}
\vskip -0.cm
\includegraphics[width=6.8cm,angle=0,clip]{fig3b.eps}
\vskip -0.cm
\hskip 0.15cm
\includegraphics[width=7.2cm,angle=0,clip]{fig3c.eps}
\caption{Energy levels of the $1s$, $1p$, $1d$, $1f$, and $1g$ $\Lambda$
single-particle shells as a function of A$^{-2/3}$, where A is the baryon number 
of the hypernucleus defined as A=N+Z+1.
Theoretical predictions are shown by bands for guiding the eye. The actual
calculations were done only for the hypernuclei listed in Table~\ref{tab:sp}.
The symbols represent experimental results \cite{gal16,Botta:2017}. 
}
\label{fig:spec}
\end{center}
\vskip -0.5cm
\end{figure}
%%%%%%%%%%%%%%%%%%%%%%%%%%%%%%%%%%%%%%%%%%%%%%%%%%%%%%%%%%%%%%%%%%%%%%

The NLO interactions, and specifically the NLO13 version, predict less bound $\Lambda$ single-particle states.
In particular, there is an underbinding of light hypernuclei such as $^5_\Lambda$He and $^{13}_\Lambda$C
while the description of the medium and heavier hypernuclei is clearly improved, cf. Table~\ref{tab:sp}.
Apparently, neither the NLO13 interaction nor the NLO19 version yield a quantitative description 
of all medium and heavy hypernuclei. 
Whereas the NLO19 interaction describes reasonably well $^{17}_{\Lambda}$O, $^{41}_{\Lambda}$Ca 
and $^{91}_{\Lambda}$Zr, it seems to overbind $^{209}_{\ \ \Lambda}$Pb. For the latter the 
predictions of the NLO13 potential are more in line with the experiment. 
We want to emphasize, however, that in view of the limitations pointed out in Sec.~\ref{sec:sec2}, 
results of the employed approach for heavy hypernuclei such as $^{209}_{\ \ \Lambda}$Pb are 
questionable. 
Indeed, the limitations are evident from Table~\ref{tab:sp} where one can see that the 
separation energy for that hypernucleus is sometimes larger than the corresponding value for 
$U_{\Lambda}(k=0)$ at nuclear matter saturation density. 

To elucidate this reduction of binding when going from LO to NLO we present in Fig.\ \ref{fig:matel}
results for selected S- and P-wave diagonal $\Lambda N$$\rightarrow$$\Lambda N$ nuclear matter
$G$-matrices as a function of the $\Lambda$-nucleon relative momentum. 
The $G$-matrices are calculated using the LO, NLO13 and NLO19 potentials, exemplary for a cutoff of $600$ MeV.
They are evaluated at nuclear matter saturation density $\rho_0=0.17$ fm$^{-3}$, zero center-of-mass momentum and a
value of the starting energy $\Omega=m_N+m_\Lambda-80$ MeV, where $-80$ MeV is an average value of the sum
of the single-particle mean fields of the nucleon ($U_N(k=k_F)\approx -50$ MeV) and the
$\Lambda$ ($U_\Lambda(k=0)\approx -30$ MeV) at this density.
We remind that the NLO13 and NLO19 potentials differ only in the S-wave channels \cite{nlo19}.
It is clear from the figure that the $\Lambda N$ $G$-matrices are, overall, less attractive for the NLO
interaction than at LO. Specifically, regarding the decisive $^1S_0$ and $^3S_1$ partial waves,
for the NLO potentials these are attractive (negative) only for small $\Lambda N$ relative
momenta $q$ but change sign, {\it i.e.} become repulsive, with increasing momentum.
Differences in the $\Lambda N$ $G$-matrices obtained with LO and NLO potentials for the higher partial waves
are found to be less than $1$ MeV\,fm$^{-3}$.
Note that the aforementioned differences in the properties of the LO and NLO interactions are 
also reflected in the corresponding $\Lambda p$ phase shifts presented in Ref.~\cite{nlo13}.

A graphic representation of our results is provided in Fig.~\ref{fig:spec}.
Here bands are shown to guide the eye, where the width
represents the variation with the cutoff \cite{nlo13,nlo19}. 
This facilitates a more direct view on the regulator dependence.
As said above, those variations can be interpreted as a lower bound for the
uncertainty of the results due to the truncation of the chiral
expansion. Note that a more realistic way for estimating this uncertainty, that does
not rely on cutoff variation, has been proposed in Refs.~\cite{Epelbaum:2015}.
However, it is not easily applicable to the present calculation.

Figure~\ref{fig:spec} unveils that the cutoff dependence of the predicted binding 
energies is sizable. Of course, considering the corresponding variation for the 
$\Lambda$ binding in infinite nuclear matter~\cite{haidenbauer15,petschauer16,nlo19}, 
it does not really come unexpectedly. Indeed, like for infinite nuclear matter, 
the variations at NLO are somewhat larger than the ones at LO -- contrary to the 
trend for $YN$ scattering in free space \cite{nlo13,nlo19} and for the binding 
energies of light hypernuclei \cite{Nogga:2013,nlo19}.
In this context we want to mention that a likewise strong regulator dependence has been 
detected in corresponding studies of nuclear matter properties in the $NN$ sector 
within chiral EFT \cite{Coraggio:2013,Sammarruca:2015,Hu:2016nkw,Hu:2019zwa}.
As discussed in Ref.~\cite{nlo19}, since the Pauli operator suppresses the contributions 
from low momenta, see Eq.~(10) of Ref.~\cite{vidana17}, the $G$-matrix results
are more sensitive to high-momentum contributions and, thus, to intermediate and 
short-distance physics \cite{Hu:2016nkw}. In the $NN$ case, indications for a convergence and a reduced 
regulator dependence were only found after going to much higher order
- N$^3$LO in Refs.~\cite{Coraggio:2013,Sammarruca:2015} and N$^4$LO in \cite{Hu:2016nkw}
- and after including three-body forces. 
Thus, we anticipate a noticeable reduction of regulator artifacts in the hyperon sector
only when 3BFs are explicitly included. This, however, is beyond the scope of the present 
work as it requires an extension of the $YN$ interaction to at least N$^2$LO \cite{nny}. 

Taking the sizable cutoff dependence into account, the predictions of the chiral 
$YN$ interactions are roughly in line with the experimental information. 
In case of the NLO19 interaction there is actually a quantitative agreement with 
the data over an extended range of $A$ values when considering the uncertainty 
due to the cutoff dependence. Only the heaviest hypernucleus studied, 
$^{209}_{\,\,\,\,\,\Lambda}$Pb, appears overbound. For the LO (NLO13) 
interactions we observe a more general tendency for overbinding (underbinding). 
Looking more carefully on the overall trend one has to conclude that the $A$ dependence 
predicted by the $YN$ interactions from chiral EFT is definitely somewhat stronger 
than the one exhibited by the data, see Fig.~\ref{fig:spec}. The very same 
trend was observed in the earlier studies of s-shell hypernuclei with phenomenological 
$YN$ interactions, within the same framework \cite{morten96,vidana98,vidana17}. 
Thus, most likely, one sees here limitations of the employed approach. 
Indeed, on the one hand, the method seems to underestimate the binding energies 
for light hypernuclei, as indicated by the LO
predictions for $^5_\Lambda$He and $^{13}_\Lambda$C in comparison to results of 
{\it ab initio} calculations~\cite{Wirth:2019}, cf. the discussion above.
On the other hand, it overestimates the energies of very heavy hypernuclei as discussed in 
Sec.~\ref{sec:sec2}. 
Of course, in principle it could be also an indication for (missing) 3BFs. Those appear 
at higher order and, therefore, are not taken into account in our calculation. Such 3BFs
are known to lead to an effective and density-dependent $\Lambda N$ interaction
\cite{Petschauer:2017}. 

%%%%%%%%%%%%%%%%%%%%%%%%%%%%%%%%%%%%%%%%%%%%%%%%%%%%%%%%%%%%%%%%%
\begin{table*}[t]
\renewcommand{\arraystretch}{1.2}
\begin{center}
\begin{tabular}{|c|cccc|c|c|}
\hline
\hline
 & \ NLO13 \ & \ NLO13+ALS \ & \ NLO19 \ & \ NLO19+ALS \ & \ NSC97e \cite{vidana17} \ & Exp. \cite{Kohri:2002} \cr
\hline 
$p_{1/2}$  & $12.43$ & $12.37$ & $13.58$  & $13.51$ & $12.63$ & $10.982\pm 0.031$ \\
$p_{3/2}$  & $12.18$ & $12.54$ & $13.22$  & $13.63$ & $12.28$ & $10.830\pm 0.031$ \\
$\Delta p$ & $0.25$  & $-0.17$ &  $0.36$  & $-0.12$ & $0.35$  & $0.152\pm 0.054\pm 0.036$ \\
\hline  
\hline
\end{tabular}
\end{center}
\caption{ Excitation energies of $^{13}_{\,\,\, \Lambda}$C (in MeV) 
for the chiral $YN$ interactions and the Nijmegen NSC97e potential. 
$\Delta p$ denotes the difference $p_{1/2}- p_{3/2}$. 
ALS indicates that an antisymmetric spin-orbit force was added to the $YN$ potential 
\cite{haidenbauer15}, see text for details. 
}
\label{tab:sp2}
\renewcommand{\arraystretch}{1.0}
\end{table*}
%%%%%%%%%%%%%%%%%%%%%%%%%%%%%%%%%%%%%%%%%%%%%%%%%%%%%%%%%%%%%%%%%

An issue often discussed in the literature is the $\Lambda$-nucleus spin-orbit interaction
where empirical information suggests that it should be rather weak \cite{gal16,botta12}. 
It is interesting to see, cf. Table~\ref{tab:sp}, that the splitting of the $p$-, $d$-, $f$- and 
$g$-wave states is of the order of few tenths of an MeV or even less in all cases, asserting indeed 
a small spin-orbit strength of the chiral $YN$ interactions. As a matter of facts, the spin-orbit
splittings for the phenomenological $YN$ potentials of the J\"ulich and Nijmegen groups 
considered in Refs.\ \cite{morten96, vidana98,vidana17} were found to be likewise small. Thus, it seems that this 
feature is well reproduced qualitatively and almost universally by interactions that 
incorporate the underlying approximate SU(3) symmetry and describe the $YN$ data. 

Very prominent cases are the splittings of the $5/2^+$ and $3/2^+$ states of $^{9}_\Lambda {\rm Be}$
and of the $3/2^-$ and $1/2^-$ states of $^{13}_{\ \Lambda} {\rm C}$ which are known
experimentally with unprecedented accuracy \cite{Akikawa:2002,Kohri:2002}. 
The ones for $^{13}_{\ \Lambda} {\rm C}$ can be calculated within our approach. 
However, it turns out that these states are only bound for the NLO interactions and
only for the cutoff $\Lambda=700$~MeV, see Table~\ref{tab:sp}.
This is already outside of the range of $500$ -- $650$~MeV considered for the NLO
interactions in Refs.~\cite{nlo13,nlo19}, guided by the achieved $\chi^2$ values.
Since the $\chi^2$ deteriorates only slightly for the larger cutoff
-- it is $17.3$ for NLO13 \cite{nlo13} and $16.5$ for NLO19 --
we believe, nonetheless, that it is instructive and sensible to take
a closer look at the predicted level splitting for the $^{13}_{\ \Lambda} {\rm C}$,
even if we have only results for the NLO interactions at $\Lambda=700$~MeV.
Certainly, in this case a direct assessment of the cutoff dependence and, accordingly,
of the uncertainty is not possible. But it is reassuring to see from the results for
heavier hypernuclei that the excitation energies, {\it i.e.} the difference between the $s$
and $p$ states, are significantly less cutoff dependend. That dependence is even further
reduced when the splitting between the $3/2^-$ and $1/2^-$ states itself is considered.

The corresponding results are summarized in Table~\ref{tab:sp2} and compared with the
BNL experiment \cite{Kohri:2002}. For the ease of comparison we include in addition 
the prediction of one of the Nijmegen NSC97 potentials, taken from Ref.~\cite{vidana17}.  
One can see that the measured excitation energies for the $3/2^-$ and $1/2^-$ states
are overestimated by about $1$ MeV. The splitting of the states is, however, reproduced
within $50$ keV by the NLO13 interaction considering the experimental uncertainty. 
The results for the NLO19 interaction and the NSC97e potential are somewhat larger
but still remarkably close to the experiment. 

In Table~\ref{tab:sp2} we show also results for a modified chiral $YN$ potential that
was introduced in Ref.~\cite{haidenbauer15}. In that work an antisymmetric
spin-orbit force (ALS) was added to the NLO interaction from Ref.~\cite{nlo13} in an 
attempt to study its influence on the in-medium properties of the $\Lambda$. It is 
generated by a contact term that facilitates $^1P_1$--$^3P_1$ transitions in the 
coupled (isospin $I=1/2$) $\Lambda N$--$\Sigma N$ system. Such a term arises 
at NLO in the Weinberg counting \cite{nlo13}, but it was simply put to zero in 
the NLO13 interaction because it could not be pinned down by a fit to the 
existing $\Lambda N$ and $\Sigma N$ scattering data. In Ref.~\cite{haidenbauer15} 
this contact term was included and the strength of the corresponding LEC was fixed by 
considering the so-called Scheerbaum factor $S_\Lambda$ \cite{SCHE} for the $\Lambda$ 
calculated in nuclear matter. The Scheerbaum factor provides a measure for the strength 
of the $\Lambda$-nuclear spin-orbit force \cite{kohno10,Fujiwara:2000} and values for it 
have been inferred, {\it e.g.,} from studies of the splitting of the $5/2^+$ and $3/2^+$ states 
of $^{9}_\Lambda {\rm Be}$ by Hiyama {\it et al.}~\cite{Hiyama:2000} and 
Fujiwara {\it et al}.~\cite{Fujiwara:2004}. 
To be concrete, the LEC in question was adjusted to yield $S_\Lambda\approx -3.7$,
cf. Ref.~\cite{haidenbauer15} for a detailed discussion of the choice, 
whereas the original NLO13 potential predicts values around $-12$. 
As can see from Table~\ref{tab:sp2}, that modification has a dramatic 
consequence for the level splitting, namely it reverses the ordering. 
Thus, the present study suggests that, for the chiral potentials NLO13 and NLO19, such 
an antisymmetric spin-orbit force has to be more moderate than the one 
introduced in \cite{haidenbauer15}, 
motivated by the studies in Refs.~\cite{kohno10,Fujiwara:2000}.  
In any case, one has to be aware that the very small level splitting results
from an extreme fine-tuning of various ingredients of the $YN$ potential. Thus, 
given the poor overall constraints on the $\Lambda N$ interaction in higher partial 
waves \cite{nlo13} and acknowleding the approximate nature of our many-body approach 
-- and of those applied in pertinent studies by other authors -- unambiguous conclusions on 
such detailed aspects of the $YN$ interaction might be difficult to draw at present.

Finally, it is worth noting that the Scheerbaum factor for the LO interaction is around $3$
\cite{haidenbauer15}, {\it i.e.,} likewise small but of opposite sign. The $p_{3/2}$ state 
of $^{13}_{\,\,\, \Lambda}$C is not bound in this case, cf. Table~\ref{tab:sp}, only the 
$p_{1/2}$ state. Notwithstanding, one can see from the results for $^{17}_{\,\,\, \Lambda}$O, say,
that the level ordering is reversed as compared to the NLO interactions. There are NCSM 
calculations of $^{9}_\Lambda {\rm Be}$ for the LO interaction~\cite{Wirth:2016,Wirth:2019}. 
They reveal that the splitting of the $5/2^+$ and $3/2^+$ 
states predicted by that interaction is indeed tiny, as suggested by the 
experiment. However, the level ordering is wrong with respect to the experiment -- 
but in line with what we observe in our own calculations for heavier hypernuclei. 

%%%%%%%%%%%%%%%%%%%%%%%%%%%%%%%%%%%%%%%%%%%%%%%%%%%%%%%%%%%%

\section{Summary and Conclusions}
\label{sec:sec4}

In this work we have studied the structure of single-$\Lambda$ hypernuclei using the chiral $YN$ potentials 
derived by the J\"{u}lich--Bonn--Munich group at LO and NLO in the chiral expansion. 
Following a perturbative many-body approach we have calculated the $\Lambda$ self-energy in finite nuclei 
from which we have determined the different $\Lambda$ single-particle bound states
in various hypernuclei from $^5_{\Lambda}$He to $^{209}_{\,\,\,\,\,\Lambda}$Pb. 
We have presented results for the LO potential \cite{lo} and the NLO 
interactions from 2013 (NLO13) \cite{nlo13} and 2019 (NLO19) \cite{nlo19}, respectively. 

It turned out that the predictions for the $YN$ interaction at LO are in rather good agreement with 
the empirical separation energies of light hypernuclei, whereas the $\Lambda$ single-particle 
energies of the other hypernuclei, specially those of $^{91}_\Lambda$Zr and $^{209}_{\,\,\,\,\,\Lambda}$Pb, 
appear clearly too large with respect to the experimental values. 
However, the good agreement for light hypernuclei is most likely questionable. Available no-core shell 
model calculations based on the same LO $YN$ interaction for $^5_{\Lambda}$He and $^{13}_{\,\,\Lambda}$C 
\cite{Wirth:2019} suggest that our approach underestimates the actual binding energies by $20$ to $30$\%
when applied to such light systems. That aspect, together with the overbinding for heavier hypernuclei,
provides a strong indication that the LO interaction is too attractive. 
 
Calculations for the NLO interactions, particularly for NLO13, yielded less bound $\Lambda$ single-particle 
states. Indeed, for NLO13 we observed an overall tendency for underbinding with respect to the empirical 
information. The predictions for NLO19 turned out to be in a qualitatively good agreement with the data 
over a fairly large range of mass number values when considering the uncertainty due to the regulator 
dependence. Only the heaviest considered hypernucleus, $^{209}_{\,\,\,\,\,\Lambda}$Pb, appears overbound
by that interaction. However, one should be rather cautious in drawing conclusions from the latter result. 
As pointed out already in Refs.\ \cite{vidana98,vidana17}, and as discussed in the present work as well, 
there are clear signals for shortcomings of the employed approach when applied to such heavy hypernuclei.
In particular, one has to keep in mind that the method itself was originally optimized for the study 
of $^{17}_{\,\,\Lambda}$O, although it is expected to work also reasonably well for, 
say, $^{13}_{\,\,\Lambda}$C and $^{41}_{\,\,\,\Lambda}$Ca. 
Independently of that, the sizable regulator dependence is a clear signal that higher-order contributions, 
and specifically three-body forces, have to play a role. 

Finally, the predicted spin-orbit splittings of the $p$-, $d$-, $f$-, and $g$-wave states are found to be very small. 
These are of the order of few tenths of MeV or even less, in remarkable agreement with the magnitude observed in 
experiments. Actually, similarly small values were also obtained in corresponding calculations \cite{vidana17} 
for meson-exchange $YN$ potentials by the J\"{u}lich and Nijmegen groups, thus, corroborating that this feature 
can be qualitatively well reproduced by interactions that take account of the underlying approximate SU(3) 
flavor symmetry and describe the $\Lambda p$ and $\Sigma N$ scattering data.

%%%%%%%%%%%%%%%%%%%%%%%%%%%%%%%%%%%%%%%%%%%%%%%%%%%%%%%%%%%%%%%

\begin{acknowledgement}
We acknowledge helpful communications with Avraham Gal, Andreas Nogga and Hirokazu Tamura. 
This work is partially supported by the COST Action CA16214 
and by the DFG and the NSFC through
funds provided to the Sino-German CRC 110 ``Symmetries and
the Emergence of Structure in QCD'' (DFG grant. no. TRR~110). 
\end{acknowledgement}

%%%%%%%%%%%%%%%%%%%%%%%%%%%%%%%%%%%%%%%%%%%%%%%%%%%%%%%%%

\end{document}